


\def\={\!=\!}
\def\-{\!-\!}
\def\a{\alpha}

\def\d{\partial}
\def\da{^{\dagger}}

\def\e{\eqno}
\def\ee{\hat\epsilon}
\def\ep{\epsilon}

\def\ie{{\it i.e.}}
\def\pg{paragrassmann}
\def\q{\quad}
\def\qq{\qquad}
\def\qm{quantum mechanics}

\def\t{\theta}

\def\Q{{\bf Q}}
\def\1{{\textstyle{1\over2}}}

\magnification=1200
\vsize=22truecm
\hsize=15truecm
\voffset=0.5truecm
\hoffset=1truecm
\baselineskip=18truept
\lineskip=1pt
\lineskiplimit=0pt
\parskip=6truept


\font\titre=cmbx10 scaled\magstep2
\font\bigletter=cmr10 scaled\magstep2


\baselineskip=12pt

\line{\hfill McGill/93-06}
\line{\hfill hep-th/9305130}
\line{\hfill May 1993}

\vskip 1in
\centerline {\titre Fractional Superspace Formulation}
\vskip 0.05in
\centerline {\titre of Generalized Mechanics}
\vskip 0.75in
\centerline{{{\bigletter S}T\'EPHANE {\bigletter D}URAND}\footnote{$^{*}$}
{E-mail address: durand@hep.physics.mcgill.ca}}
\vskip 0.2in
\centerline{\it Department of Physics}
\centerline{\it McGill University}
\centerline{\it 3600 University Street}
\centerline{\it Montr\'eal, PQ, H3A 2T8, Canada}
\vskip 0.45in

\centerline{To appear in {\it Mod. Phys. Lett.} {\bf A8}, No.24 (1993).}
\vskip 0.45in

\centerline{\bf Abstract}
\vskip 0.1in

\noindent
Supersymmetric (pseudo-classical) mechanics has recently been generalized to
{\it fractional}\/ supersymmetric mechanics. In such a construction, the action
is invariant under fractional supersymmetry transformations, which are the
$F^{\,\rm th}$ roots of time translations (with $F=1,2,...$).
Associated with these symmetries, there are conserved charges
with fractional canonical dimension $1+1/F$.
Using \pg\ variables satisfying $\t^F=0$, we present a fractional-superspace
formulation of this construction.

\vfill
\eject

\baselineskip=18truept

\noindent {\bf 1. Introduction}
\vskip 0.2cm

In field theory, supersymmetric (SUSY) transformations are ``square
roots" of {\it space-time}\/ translations. As a special case, (quantum)
mechanics can be considered as a
field theory in one time and zero space dimensions, and in this context, SUSY
transformations are then the square roots of {\it time}\/ translations
only. Since time translations are generated by the Hamiltonian $H$, and SUSY
transformations by the supercharge $Q$, we are led to the
algebra $H=Q^2$ which defines SUSY \qm.

We recently have presented a new generalization of SUSY \qm\ which we call
{\it fractional}\/ supersymmetric (FSUSY) \qm.$^{[1,2]}$
In such a construction, the Hamiltonian
is expressed as the
$F^{\,\rm th}$ power of a conserved {\it fractional} supercharge: $H=Q^F$, with
$[H,Q]=0$ and $F=1,2,...\, .$ Motivated by this new algebra,
we have {\it pseudo-classically}\/ described FSUSY transformations as the
$F^{\,\rm th}$ roots of time translations, and provided an action invariant
under such transformations.$^{[1,3]}$ Here, we shall reformulate these results
in fractional superspace, using \pg\ variables of order $F$ satisfying
$\t^F=0$. Additionally, we
construct the conserved N\"other fractional supercharges, which are of
fractional canonical dimension $1+1/F$, associated with these symmetry
transformations.
In the present work, we restrict
ourselves to the free particle. The interacting case will be discussed
elsewhere.$^{[4]}$

\vskip 0.5cm
\noindent {\bf 2. Fractional superspace formalism}
\vskip 0.2cm

In this section, we introduce paragrassmann variables which
interpolate between ordinary bosonic and fermionic variables, and which
will be used to construct fractional-superspace coordinates. (In a
quantum-mechanical context, these \pg\ variables are instead interpreted as
generalized creation and annihilation operators.$^{[1,2]}$)
Note that we will use the following definition:
$$[A,B]_\omega\equiv AB-\omega BA. \eqno(1)$$

We introduce a {\it real}\/ paragrassmann variable $\t$ of order $F$, and
its (real) derivative $\d\equiv \d/\d\t$, which satisfy
$$\eqalign{\t^F=0&, \qq \d^F=0, \qq F=1,2,...   \cr
          &(\t^{F-1}\not=0, \; \d^{F-1}\not=0). \cr} \eqno(2)$$
We take the
generalized commutation relation between $\t$ and $\d$ to be
$$[\d,\t]_q=\a(1-q)         \eqno(3)$$
where $\a$ is an arbitrary {\it real}\/ parameter, and $q\in\cal C$ a {\it
primitive}\/
$F^{\,\rm th}$ root of unity:
$$q^F=1 \qq (q^n\not=1 ~~{\rm for}~~ 0<n<F).  \eqno(4)$$
By a {\it primitive}\/ root, we mean a root satifying the
condition in parentheses; for instance, $q\not=\pm 1$ for $F=4$.
We require the factor $\a(1-q)$ with a real (but arbitrary) $\a$ in the r.h.s
of (3) so that (3) is consistent both under complex conjugation, and in the
``null" case $F=1$ ($q=1$), \ie, $\t=\d=0$.
Note that we recover
the ordinary grassmann case for $F=2$ $(q=-1)$: $\t^2=\d^2=0$ and
$\{\d,\t\}=2\a$.
Moreover, for some choices of $\a$ (for instance, $\a=F$),
we also recover the bosonic case in the limit $F\to\infty$ $(q=1)$:
$[\d,\t]={\it const}$.$^{[1,2]}$
However, in the following sections, we will not be concerned with this limit,
so we let $\alpha$ remain unfixed.
The definition (3) implies
$$\d\cdot\t^n=\a(1-q^n)\,\t^{n-1}+q^n\t^n\d. \eqno(5)$$
Setting $n=F$ in $(5)$ demonstrates that the condition $(4)$ is actually a
consequence of $(2)$ and $(3)$.

For a given order $F$ which is $prime$, we can actually introduce $F-2$ other
fractional derivatives. We thus have $F-1$ derivatives, which we write as
$\d_i$ with
$i=1,2,...,F-1$. With this notation, $\d=\d_1$. These derivatives satisfy
$$\d_i^F=0, \qq \d_i^{F-1}\not=0, \qq
[\d_i,\t^n]_{q^{i\!n}}=\a(1-q^{in})\t^{n-1} \eqno(6)$$
and have the properties
$$[\d_i,\d_{F-i}]_{q^{-i}}=0. \eqno(7)$$
Note that in the SUSY limit $(F=2;q=-1)$ we are left with only one derivative
satisfying the usual relation: $[\d,\d]_{q^{-1}}=\{\d,\d\}=2\,\d^2=0$. For a
non-prime $F$, the situation is more complicated.
For instance, for $F=4$, setting $i=2$ in (7) implies $(\d_2)^2=0$, which
contradicts (6). In other words, $q^i$ is not a primitive root for $i=2$.
Therefore, for a non-prime $F$, there are fewer than $F\-1$ derivatives $\d_i$,
but at least two: $\d_{\t}\equiv\d_1$ and $\delta_{\t}\equiv\d_{F-1}$.
(We now put a $\t$-subscript on the $\t$-derivatives to clearly distinguish
them from the time-derivative $\d_t\equiv \d/\d t$ and from the variation
$\delta$.)
In the following, we will use only these two derivatives.

We now introduce the {\it fractional covariant}\/ derivative
$$D=\delta_\t+ie\t^{F-1}\d_t, \qq D^F=i\d_t                         \eqno(8)$$
and the generator of FSUSY transformations
$$Q=\d_\t-ie\t^{F-1}\d_t, \qq Q^F=-i\d_t                         \eqno(9)$$
where $e^{-1}=F\a^{F-1}$. These satisfy
$$[D,Q]_q=0.  \e(10)$$
Note that in the SUSY case $(F=2)$, we have $\delta_\t=\d_\t$ and $\{Q,D\}=0$.
Note also that the null limit $\d_\t=0$ is needed for the consistency of (8)
and (9) when $F=1$.

Finally, we define the integration over a (real) paragrassmann variable $\t$ of
order $F$ as
$$\int d\t\;\t^p=\delta_{p,F-1}.            \eqno(11)$$
This direct generalization of integration over ordinary grassmann
variables is invariant under translation ($\t\to\t+\epsilon$), as in the
grassmann case. Note however that
unlike the grassmann case,
the results of derivation and integration are different. Also note that there
is no bosonic limit of such a definition. This is not a problem since we will
not require the bosonic limit of a \pg\ variable, but rather the bosonic limit
of the fractional-superspace coordinate (which will correspond to the case
$F=1$).

\vskip 0.5cm
\noindent {\bf 3. Ordinary SUSY mechanics}
\vskip 0.2cm

Let us first recall the ordinary SUSY case. We work in one dimension. The
position of the particle is described by $x(t)$,
whereas its internal space is described
by a {\it real}\/ fermionic variable
$\psi(t)$ satisfying $\psi^2=0$. The SUSY transformations are given by
$$\eqalignno{\delta x&=i\epsilon\,\psi &(12a)\cr
             \delta\psi&=\epsilon\,\dot x &(12b)\cr}$$
where $\epsilon$ is a real fermionic infinitesimal parameter. Since
$\epsilon\psi=-\psi\epsilon$, the r.h.s of Eqs.~(12) are real. Note that
$\delta^2 x=i\epsilon_1\epsilon_2\,\dot x$ and $\delta^2
\psi=i\epsilon_1\epsilon_2\,\dot\psi$, so SUSY transformations are indeed the
square roots of time translations. An action invariant under (12) is
$$S=\int dt\;{1\over2}(\dot x^2+i\dot\psi\psi). \e(13)$$
More precisely, the Lagrangian of (13) varies by the total time derivative
$\d_t({i\over2}\dot x\psi)$. Note that since $\psi\dot\psi=-\dot\psi\psi$, this
action is real.

To give a superspace formulation of this construction, we need a real grassmann
coordinate $\t$ and its derivative $\d_\t$ which satisfy
$$\t^2=\d_\t^2=0, \qq \{\d_\t,\t\}=1; \e(14)$$
these are the relations (2-4) for $F=2$ (and $\a=1/2$). We also recall
the rule of integration over ordinary grassmann variables:
$$\int d\t\;(a+\t\,b)=b. \eqno(15)$$
Then, we combine the
variables $x$ and $\psi$ into a superspace coordinate $Z(t,\t)$:
$$Z(t,\t)=x(t)+i\t\,\psi(t). \e(16)$$
We say that $x$ and $\psi$ respectively belong to the sector-0 and sector-1 of
the theory ($\t$ is also in sector-1 and sectors are defined modulo 2). Note
that $Z$ is real since $\psi\t=-\t\psi$. In this formalism, the SUSY
transformations are generated by
$$Q=\d_\t-i \t\d_t,\qq  Q^2=-i\d_t.  \eqno(17)$$
Acting on $Z(t,\t)$, we have
$$\delta Z=\epsilon QZ                   \eqno(18)$$
which gives in components the transformations (12).
We also need the covariant derivative
$$D=\d_\t+i\t\d_t,\qq  D^2=i\d_t  \eqno(19)$$
which anticommutes with $Q$:
$$\{D,Q\}=0. \e(20)$$
The relation (20) implies that if $A$ transforms as $\delta A=\epsilon QA$
(\ie, if $A$ is a superfield), then $DA$ transforms in the same way,
$\delta(DA)=D(\delta A)=D(\ep QA)=\ep Q(DA)$, since $\ep D=-D\ep$ (\ie, since
$\ep$ anticommutes with $\t$ and $\d_\t$). Thus, the covariant derivative of a
superfield is again a superfield. Moreover, the product of two superfields is
also a superfield since the Leibniz rule holds:
$$\ep Q(AB)=\ep Q(A)\,B+A\,\ep Q(B). \e(21)$$
Indeed, we have $\delta(AB)=(\delta A)B+A(\delta B)=(\ep QA)B+A(\ep QB)=\ep Q
(AB)$. Therefore, it is easy to construct an invariant action. For instance,
the action
$$S=-\int dt\,d\t\;{1\over2}(D^2\! Z\,D\! Z)
 =-i\int dt\,d\t\;{1\over2}(\dot Z\,D\! Z)      \eqno(22)$$
is invariant under (18) since
$\delta S=-{1\over2}\int dt\,d\t\,\ep Q(D^2\! Z\,D\! Z)$, which is
automatically zero (as we will see in Sect.~5 for the general case). Using the
rule (15) to integrate over $\t$, we find the action (13).

\vskip 0.5cm
\noindent {\bf 4. FSUSY mechanics of order 3}
\vskip 0.2cm

Let us now turn to the next order: we now want to construct an action invariant
under {\it cube}\/ roots of time translations.
We first introduce
the fractional-superspace coordinate $Z(t,\t)$ of order 3:\footnote{$^\dagger$}
{Throughout this paper, we will use the following convention: for $q=e^{i\t}$
with $0\leq\t<2\pi$, the square root is defined as $q^{1/2}=e^{i\t/2}$.}
$$Z(t,\t)=x(t)+q^{1/2}\,\t\,\psi(t)+q^2\,\t^2\phi(t)          \eqno(23)$$
where $\t$ is a {\it real}\/ \pg\ variable satisfying $\t^3=0$, and
where $q$ is a primitive cube root of unity, \ie, one which satisfies $q^3=1$
with $q\not=1$. The factors $q^{1/2}$ and $q^2$ in (23) are needed in order for
$Z$ to be real (see below). We thus have a theory with three sectors, and
sectors are defined modulo 3.
The bosonic variable $x(t)$ remains a sector-0 quantity, but now we have
{\it two}\/ types of {\it real}\/ internal-space variables, $\phi(t)$ and
$\psi(t)$, which respectively belong to sector-1 and sector-2 since we take
$\t$ to be in sector-1 (and since $Z(t,\t)$ must be a sector-0 quantity).
We take the following commutation relations between the new variables:
$$\eqalignno{&\t\phi=q\phi\t, \qq \t\psi=q^2\psi\t  &(24a)\cr
\noalign{\hbox{and}}
             &\psi\dot\phi=q\dot\phi\psi, \qq \phi\dot\psi=q^2\dot\psi\phi.
&(24b)\cr}$$
We now can see that $Z^{\ast}=Z$.
Since we have chosen to work with a real paragrassmann variable $\t$, the
series in $(23)$ contains only 3 terms and no auxiliary field. [Alternatively,
we could have chosen to work with a complex $\t$, we then would have had a
fractional-superspace coordinate $Z(t,\t,\t^{\ast})$ containing 9 terms, among
which some would have been auxiliary fields. But here we want to present the
construction in the simpler formulation of (23).]

The FSUSY transformations of order 3 are the cube roots of a time translation.
They are generated by the FSUSY generator
$$Q=\d_\t-i e\,\t^2\d_t,\qq  Q^3=-i\d_t  \eqno(25)$$
where $e^{-1}=3\a^2$, and where $\t$ and $\d_\t$ satisfy (2-4) for $F=3$.
Acting on $Z(t,\t)$, we have
$$\delta Z=(iq^{-1/2})\,\epsilon QZ                   \eqno(26)$$
which gives in components
$$\eqalignno{&\delta x  =i\epsilon\,\a(1-q)\,\psi       &(27a)\cr
             &\delta\psi=i\epsilon\,\a(1-q^2)\,\phi  &(27b)\cr
             &\delta\phi=\epsilon\,e\,\dot x  &(27c)\cr}$$
where $\epsilon$ is a real sector-1 infinitesimal parameter,
which must satisfies\footnote{$^\dagger$}
{To pass from (26) to (27), we need to use either $(24a)$ or (28).}
$$\ep\t=q^{-1}\t\ep, \qq \ep\d_\t=q\d_\t\ep, \qq \ep\delta_\t=q\delta_\t\ep.
\e(28)$$
In order for the r.h.s of (27) to be real, $\ep$ must satisfy the following
commutation relations with the different fields:
$$\ep x=x\ep, \qq \epsilon\phi=q\phi\epsilon, \qq
\epsilon\psi=q^2\psi\epsilon. \e(29)$$
We easily see that $\delta^3=-\epsilon_1\epsilon_2\epsilon_3\,\d_t$ and thus
Eqs.~(27) indeed represent cube roots of time translations.
An action invariant under (27) is
$$S=\int dt\;{1\over2}\big[{\dot x}^2+i\beta(1-q)\dot\phi\psi
+i\beta(1-q^2)\dot\psi\phi\big] \eqno(30)$$
with $\beta=3\a^3$.
This Lagrangian is real according to $(24b)$, and varies by a total derivative
under the transformations (27) [see below]. In Sect.~6, we will construct the
associated conserved fractional charge.

Let us  now write the action (30) in a fractional-superspace formulation, in
order to make its FSUSY invariance manifest. First, we define integration over
a paragrassmann variable of order 3 as
$$\int d\t\;(a+\t\,b+\t^2 c)=c \eqno(31)$$
which is the particular case $F=3$ of (11). Then, we introduce the fractional
covariant derivative of order 3
$$D=\delta_\t+i e\,\t^2\d_t,\qq  D^3=i\d_t  \eqno(32)$$
which satisfies the commutation relation (10) with $Q$. For $D\! Z$ to
transform in the same way as $Z$ in (26), we must have $[\ep Q,D]=0$, \ie\
$\ep D=qD\ep$, which follows from (28).
Moreover, according to $(24a)$, (28) and (29), we can show that the Leibniz
rule (21) still holds.\footnote{$\da$}
{The Leibniz rule works differently for $D$ and $Q$, as we will see in
Sect.~6.}
Hence, the fractional covariant derivative of a fractional superfield is also a
fractional superfield, and the product of two fractional superfields is again a
fractional superfield. For instance, the action
$$S=-\int dt\,d\t\;{1\over2e}(D^3\! Z\,D\! Z)
 =-i\int dt\,d\t\;{1\over2e}(\dot Z\,D\! Z)      \eqno(33)$$
is invariant under (26) since the variation of the Lagrangian can be written as
a
total $Q$-derivative, which implies $\delta S=0$ (as we will see in Sect.~5).
Integrating out the paragrassmann variable $\t$ via the rule (31), we obtain
the action (30).

\vskip 0.5cm
\noindent {\bf 5. FSUSY mechanics of arbitrary order}
\vskip 0.2cm

Let us now turn to the general case, \ie, the $F^{\,\rm th}$ roots of time
translations with $F=1,2,...\,$. We introduce the fractional-superspace
coordinate $Z(t,\t)$ of order $F$:
$$Z(t,\t)=\sum_{i=0}^{F-1}c_i\,\t^i\psi_{(i)}=x(t)+
\sum_{i=1}^{F-1}c_i\,\t^i\psi_{(i)}, \qq c_i=q^{i^2/2} \eqno(34)$$
where $\t$ is a real \pg\ variable satisfying $\t^F=0$,
where $\psi_{(0)}\equiv x(t)$, and where $q$ is a primitive $F^{\,\rm th}$ root
of unity [\ie, satisfies (4)].
Since we take $\t$ to be in sector-1, the {\it real}\/ fields
$\psi_{(i)}=\psi_{(i)}(t)$ belong to the sector-($F-i$). Sectors are defined
modulo $F$.
Note that the case $F=1$ corresponds to an ordinary bosonic variable.
We introduce the following commutation relations:
$$\eqalignno{\t\,\psi_{(i)}&=q^{-i}\,\psi_{(i)}\,\t  &(35a)\cr
  \psi_{(i)}\psi_{(F-i)}&=q^i\,\psi_{(F-i)}\psi_{(i)}. &(35b)\cr}$$
The relation $(35a)$ implies that $Z^{\ast}=Z$. Consistent with $(35b)$, we
also introduce the commutation relation
$$\psi_{(i)}\dot\psi_{(F-i)}=q^i\,\dot\psi_{(F-i)}\psi_{(i)}. \e(36)$$
For $F=2$, the relations $(35)$ and $(36)$ reduce to the proper SUSY results
$\t\psi=-\psi\t$, $\psi^2=0$ and
$\psi\dot\psi=-\dot\psi\psi$ [where $\psi\equiv\psi_{(1)}$]. With the
identification $\psi_{(F)}\equiv x$, the relation (36) also implies $\dot x
x=x\dot x$.

The FSUSY transformations of order $F$, \ie\ the $F^{\,\rm th}$ roots of time
translations, are generated by the FSUSY generator $Q$ given in (9).
It  belongs to the sector-($F-1$) since $\t$ and $\d_\t$ belong respectively to
the sectors 1 and  $F-1$.
The variation $(26)$ now gives
in components $(i=1,2,...,F-1)$:
$$\eqalignno{&\delta\psi_{(i-1)}=i\epsilon\,\a(1-q^i)\,\psi_{(i)}&(37a)\cr
             &\delta\psi_{(F-1)}=\epsilon\,e\,\dot x  &(37b)\cr}$$
where the real infinitesimal parameter $\epsilon$ belongs as before to the
sector-1, and satisfies (28).  We now have $\delta^F
\psi_{(i)}=i^{F-1}\epsilon_1...\epsilon_F\,\dot\psi_{(i)}$ [since
$\prod_{i=1}^{F-1}(1-q^i)=F$], and thus the transformations (37) are indeed
$F^{\,\rm th}$ roots of time translations.
To ensure that the r.h.s of the transformations $(37)$ are real (and that the
action given below is invariant), we must take the following commutation
relations between $\epsilon$ and $\psi_{(i)}$:
$$\epsilon\,\psi_{(i)}=q^{-i}\psi_{(i)}\epsilon.            \e(38)$$
For later convenience, we may combine the equations $(37)$ as
$(i=0,1,...,F-1)$:
$$\delta\psi_{(i)}=i\epsilon\a(1-q^{i+1})\,\psi_{(i+1)}+
  \epsilon e\,\dot x\,\delta_{i,F-1}.  \e(39)$$
Our previous results (12) and (27) are recovered
for $F=2$ and $F=3$, with the notation $\psi_{(1)}\equiv\psi$ and
$\psi_{(2)}\equiv\phi$. For $F=1$, we simply have $\delta x=\epsilon\,\dot x$.
Note that in (26), $iq^{-1/2}=1$ for $F=2$.
An action invariant under (39) is:
$$S=\int dt\;{1\over2}\,\big[{\dot
x}^2+i\beta\sum_{i=0}^{F-1}(1-q^{-i})\dot\psi_{(i)}\psi_{(F-i)}\big] \e(40a)$$
with $\beta=\a e^{-1}=F\a^F$. More precisely, the Lagrangian varies by the
total derivative
$$\delta L={d\over dt}\,{\textstyle{\epsilon\over2}}\!\!\left[i\a(1-q)\,\dot
x\,\psi_{(1)}+\dot x^2\,\delta_{F,1}\right]\equiv {d\over dt}\,X. \e(41)$$
Alternatively, using
$(36)$ and the symmetry of the action under the substitution $i\to F-i$, we may
rewrite $(40a)$ as
$$S=\int dt\;{1\over2}\,\big[{\dot
x}^2-i\beta\sum_{i=0}^{F-1}(1-q^{-i})\psi_{(i)}\dot\psi_{(F-i)}\big]. \e(40b)$$
Note that the action is real.
The first term of the sum in $(40)$ always vanishes, but is included because it
allows us to take the $F=1$ limit, which corresponds to the
spinless (free) particle:
$$S=\int dt\,{1\over2}{\dot x}^2. \e(42)$$
The cases $F=2$ (with $\a=1/2$) and $F=3$ reduce to those given previously in
(13) and (30).

The fractional-superspace formulation of the action is:
$$S=-\int dt\,d\t\;{1\over2e}\,(D^F\! Z\,D\! Z)
 =-i\int dt\,d\t\;{1\over2e}\,(\dot Z\,D\! Z)      \eqno(43)$$
where $D$ is given in (8). In such a formulation, a general action is invariant
under a transformation $Z\to Z+\delta Z$ if the variation of the Lagrangian can
be written as a total derivative,
$$\delta L=\d_t\Lambda_1+\d_{\t}\Lambda_2+\delta_{\t}\Lambda_3,  \e(44)$$
since the time-integrations of $\d_t\Lambda_1$ and the $\t$-integrations of
$\d_{\t}\Lambda_2$ and $\delta_{\t}\Lambda_3$ all vanish (there is no
$\t^{F-1}$ term in $\d_{\t}\Lambda_2$ and $\delta_{\t}\Lambda_3$ which would
have survived the $\t$-integration). In particular, a variation of the
Lagrangian which is a total $Q$-derivative or $D$-derivative,
$$\delta L=Q(\Lambda)+D(\Lambda'),  \e(45)$$
can be written as in (44).
Therefore, the action (43) is automatically invariant under the variation (26)
since $\delta(\dot Z D\!Z)=\ep Q(\dot Z D\!Z)$; to see this last point, recall
the discussion between formulas (32) and (33), replacing $(24a)$ and (29) by
$(35a)$ and (38).
Using the rule (11) to integrate over $\t$ in (43), we obtain the action (40).
In order to recover the action of a spinless particle (42) directly from (43)
for $F=1$, we need to define the integration over a \pg\ variable of order 1 as
$\int d\t=1$, \ie, as the $F=1$ case of the formula (11).

\vskip 0.5cm
\noindent {\bf 6. Fractional supercharges and Euler-Lagrange equations}
\vskip 0.2cm

We first introduce the generalized momenta conjugate to $\psi_{(i)}$:
$$\eqalignno{&\pi_{(0)}\equiv{\d L\over\d\dot x}=\dot x   &(46a)\cr
             &\pi_{(i)}\equiv2\,{\d L\over\d\dot\psi_{(i)}}
     =i\beta(1-q^{-i})\,\psi_{(F-i)}, \qq i=1,...,F-1     &(46b)\cr}$$
(the factor 2 will be discussed below).
They are the components of the fractional-superspace momentum conjugate to
$Z(t,\t)$
$$\Pi(t,\t)\equiv 2\,{\d L\over\d\dot Z}=-{i\over e}\,D\!Z  \e(47)$$
which is decomposed as
$$\Pi(t,\t)=\sum_{i=0}^{F-1}a_i\,\t^i\pi_{(F-1-i)},
\qq a_i=q^{(i^2-1)/2}. \e(48)$$
Note that $\Pi^{\ast}=\Pi$. In order to recover the proper result for $F=1$,
i.e. $p=\dot x$, we must consider $\dot Z$ and $D\!Z$ as independent variables
when calculating (47) even for $F=1$.
The action is then simply
$$S=\int dt\,d\t\;{1\over2}(\dot Z\,\Pi).  \e(49)$$
We also have
$$\int d\t\;[\,\Pi\,,Z\,]=0 \qq {\rm and} \qq \int d\t\;[\,\Pi\,,\dot Z\,]=0
\e(50)$$
which follow from (35). If we wish to add
$$[\,\dot Z\,,Z\,]=0 \e(51)$$
we must require
$$\psi_{(i)}\dot\psi_{(j)}=\dot\psi_{(j)}\psi_{(i)} \q {\rm for} \q j\not=F-i.
\e(52)$$

We now wish to construct the conserved N\"other charges associated with the
symmetry transformations $(39)$. It is important to notice that for a general
Lagrangian of the form
$L=\sum_{i=0}^{F-1}\a_iA_{(i)}B_{(F-i)}$ with
$A_{(i)}B_{(F-i)}=q^{\pm i}B_{(F-i)}A_{(i)}$ (we concentrate here on the
internal-space part), we have
$$\delta L\equiv
\sum_{i=0}^{F-1}\a_i\big[\delta A_{(i)}B_{(F-i)}+A_{(i)}\delta B_{(F-i)}\big]
=\sum_{i=0}^{F-1}\big[\delta A_{(i)}{\d L\over\d A_{(i)}}+
\delta B_{(i)}{\d L\over\d B_{(i)}}\big].   \e(53)$$
Then, it is easy to show that the generalized Euler-Lagrange equations which
follow from a least-action principle are
$${\d L\over\psi_{(i)}}-{d\over dt}\bigg({\d L\over\dot\psi_{(i)}}\bigg)=0.
\e(54)$$
Therefore,
the quantity $$C=\sum_{i=0}^{F-1}\delta\psi_{(i)}\,{\d
L\over\d\dot\psi_{(i)}}\,-\,X, \qq {dC\over dt}=0 \e(55)$$
is a constant of motion when the Lagrangian varies under a transformation
$\delta\psi_{(i)}$ by the total derivative $\delta L=dX/dt$. The Hamiltonian is
simply the particular case $\delta\psi_{(i)}=\dot\psi_{(i)}$:
$$H=\sum_{i=0}^{F-1}\dot\psi_{(i)}\,{\d L\over\d\dot\psi_{(i)}}\,-\,L
\,=\,{1\over2}\,{\dot x}^2. \e(56)$$
For $\delta\psi_{(i)}$ given by $(39)$ and corresponding $X$ given in $(41)$,
we find the following conserved fractional supercharge:\footnote{$\da$}
{If we had defined $\pi_{(i)}$ in $(46b)$ without the factor 2, the boundaries
of the summation in (57) would have been $\sum_{i=1}^{F-1}$, whereupon the case
$F=1$ would not have been included. Moreover, the definitions (46) are those
which follow naturally from the fractional-superspace formulation (47).}
$$\Q={1\over2}\sum_{i=0}^{F-1}\pi_{(F-1-i)}\pi_{(i)}. \e(57)$$
The three first orders are explicitly:
$$\eqalignno{&F=1: \qq \Q=\1 \dot x^2 &(58a)\cr
             &F=2: \qq \Q=a\,\dot x\psi &(58b)\cr
             &F=3: \qq \Q=b\,\dot x\psi+c\,\phi^2 &(58c)\cr}$$
where
$$a=2i\beta, \qq b=i\beta(1-q), \qq c=\textstyle{3\over2}\beta^2q^2.\e(59)$$
Note that $(\ep\Q)^{\ast}=\ep\Q$, \ie, $\Q^{\ast}=q^{-1}\Q$, using (52);
however, $q^{-1/2}\Q$ is a real charge. For $F=1$, $\Q$ is simply the
Hamiltonian, and for $F=2$ we recover the usual SUSY charge.
The canonical dimension of $\Q$ is $\Delta[\Q]=1+1/F$ (we have
$\Delta[\d_t]=1$, $\Delta[\t]=-1/F$ and $\Delta[\psi_{(i)}]=i/F$).
It would be natural to guess that other FSUSY transformations are generated by
different powers of $Q$, \ie, $Q^k$ with $k=1,...,F$. However, if one tries to
apply the Leibniz rule (21) with $\ep'Q^k$, one finds that it holds only for
$k=1$ and $k=F$, \ie, for $Q$ and $\d_t$. Therefore, there are conserved
charges with dimension $\Delta=1+k/F$ only for $k=1$ and $k=F$. Note also that
the charge $\Q$ can be seen as a ``one time and zero space" version of a
conserved fractional current. On the other hand, the generator $Q$ of FSUSY
transformations is one element of an infinite algebra which we call {\it
fractional super-Virasoro algebra}.$^{[5]}$ In this context, $Q$ has fractional
``conformal spin" $s=1+1/F$.

Finally, let us write the fractional-superspace formulation of the
Euler-Lagran-ge equation of motions. We first emphasize that in order for the
operator $\ee D$ to obey the Leibniz rule, as $\ep Q$ in (21), the parameter
$\ee$ must be different from $\ep$ [which satisfies (28)]. Rather, we have
$$\ee D(AB)=\ee D(A)\,B+A\,\ee D(B) \e(60)$$
with
$$\ee\t=q\t\ee, \qq \ee\d_\t=q^{-1}\d_\t\ee \qq
\ee\delta_\t=q^{-1}\delta_\t\ee. \e(61)$$
Unlike $\ep$, this new parameter $\ee$ commutes with $Z$ and $\dot Z$.
Therefore, in the particular case where $A$ is a function of $Z$ and $\dot Z$
only, \ie\ $A\not=A(D\!Z)$, we have simply
$$D(AB)=D(A)B+A\,D(B), \qq A=A(Z,\dot Z). \e(62)$$
As we will see below, we must restrict ourself to Lagrangians of the form
$L=L(Z,D\!Z,\dot Z)$, \ie, with no dependence on terms as $D^i\!Z$ with
$1<i<F$.
Then, a general variation of the action is written as
$$\delta S=\int dt\,d\t\,\Big(\delta Z\,{\d L\over\d Z}+\delta D\!Z\,{\d
L\over\d D\!Z}+\delta\dot Z\,{\d L\over\d\dot Z}\Big). \e(63)$$
In order to write $\delta S$ such as in (63), all the variables must commute
under the $\t$-integration, \ie, we need (50) and (51). However, in general
there is no term such as $\dot Z Z$, either in free part (49) or in the
interaction part (see below), so we can forget (51).
As usual, the last term is rewritten as $-\delta Z\,\d_t(\d L/\d\dot Z)$.
Using (62) and $\int dt\,d\t\,D(\Lambda)=0$, we can rewrite the second term as
$-\delta Z\,D(\d L/\d D\!Z)$.
Then, requiring $\delta S=0$ for arbitrary variations $\delta Z$ leads to
the following Euler-Lagrange equation:
$${\d L\over\d Z}-D\bigg({\d L\over\d D\!Z}\bigg)-{d\over dt}\bigg({\d L\over\d
\dot Z}\bigg)=0. \e(64)$$
Note that, as mentioned before, $\dot Z$ and $D\!Z$ must be considered as
independent variables even for $F=1$.
We now see why we must consider only Lagragians of the form $L=L(Z,D\!Z,\dot
Z)$, since the least-action principle yields the equation of motion only
through the Leibniz rule. The formula (64) is a generalization of the
superspace Euler-Lagrange equation of Ref.~[6].

For the present free-particle case, the equation of motion is trivial. Indeed,
from (43) and (64), we obtain $\dot\Pi=0$, \ie, $\ddot x=0$ and
$\dot\psi_{(i)}=0$ for $i>0$. However, recalling that $Z$, $D\!Z$ and $\dot Z$
are fractional superfields and that any product of such fields is again a
fractional superfield, it is straightforward to add magnetic interactions which
preserve the FSUSY invariance.$^{[4]}$ For instance (in more than one
dimension), a term such as $A_i(Z)D\!Z_i$ leads, after integration over $\t$,
to $L_{\rm int}=A_i(x)\dot x_i+$ spin-magnetic-field couplings.

\vskip 0.5cm
\noindent {\bf 7. Concluding remarks}
\vskip 0.2cm

As mentioned in the introduction, the relation $Q^F=-i\d_t$ is inspired from
the FSUSY \qm\ ${\hat Q}^F=\hat H$, where ${\hat Q}$ is the {\it quantum}\/
fractional supercharge and $\hat H$ the Hamiltonian.$^{[1,2]}$ However, there
is another generalization of SUSY \qm\ called para-supersymmetric (PSUSY) \qm
$^{[7]}$. For instance, PSUSY \qm\ of order 3 is defined through the relation
${\hat Q}^3={\hat Q}\hat H$ with ${\hat Q}^2\not=\hat H$.
Correspondingly, there is a generator of PSUSY transformations,
$Q=\d_\t-ie\t\d_t$ with $\t$ a \pg\ variable of order 3 (and $e^{-1}=3\a$),
which satisfies $Q^3=-iQ\d t$ (but $Q^2\not=-i\d t$).$^{[5]}$ The
para-superspace coordinate is of the form $Z(t,\t)=x+\t\psi+\t^2B$, where $B$
is a {\it bosonic}\/ auxiliary field since the theory has only two
sectors.$^{[4]}$


\vskip 0.5cm
\centerline{\bf Acknowledgments}
\vskip 0.2cm

I am pleased to thank Keith Dienes for
very valuable discussions and comments.
This work is supported in part by a fellowship from
the Natural Sciences and
Engineering Research Council (NSERC) of Canada.


\vskip 0.5cm
\centerline{\bf References}
\vskip 0.2cm
\par
\frenchspacing

\item{[1]}
S. Durand, ``Fractional Supersymmetry and Quantum Mechanics",
preprint McGill/92-54 (April 1993) hep-th/9305128,
to appear in {\it Phys. Lett.} {\bf B312}, 115 (1993).

\item{[2]}
S. Durand, ``Extended Fractional Supersymmetric Quantum Mechanics",
pre-print McGill/93-03 (April 1993) hep-th/9305129,
to appear in {\it Mod. Phys. Lett.} {\bf A8}, 1795 (1993).

\item{[3]}
S. Durand, ``Fractional Supersymmetry",
preprint McGill/93-05 (March 1993) to appear in the {\it Proceedings of the
VII J.A. Swieca Summer School on Particles and Fields},
Campos do Jord\~ao, Brazil, January 1993 (World Scientific, 1993).

\item{[4]}
S. Durand, in preparation.

\item{[5]}
S. Durand, {\it Mod. Phys. Lett.} {\bf A7}, 2905 (1992).

\item{[6]}
E. D'Hoker and L. Vinet, {\it Lett. Math. Phys.} {\bf 8}, 439 (1984).

\item{[7]}
V.A. Rubakov and V.P. Spiridonov, {\it Mod. Phys. Lett.} {\bf A3}, 1337 (1988);
S. Durand, M. Mayrand, V.P. Spiridonov and L. Vinet, {\it Mod. Phys. Lett.}
{\bf A6}, 3163 (1991).

\par
\vfill
\end